\nonstopmode

\documentclass[11pt]{article}

\usepackage[fleqn]{amsmath}

\usepackage{hyperref}
\hypersetup{
    colorlinks=true,
    linkcolor=blue,
    filecolor=magenta,      
    urlcolor=blue,
}

\usepackage{tcolorbox}   
\definecolor{mycolor}{rgb}{0.122, 0.435, 0.698}

  \usepackage{geometry}
\usepackage{amsmath,amssymb,amsthm}
\usepackage{graphics,epsfig,calc}

\usepackage{latexsym,epsfig,bm,amssymb}
\usepackage{xcolor}
\usepackage{amsthm,mathrsfs}

\usepackage{mathptmx}

\DeclareSymbolFont{AMSb}{U}{msb}m{n}
\DeclareSymbolFontAlphabet{\mathbb}{AMSb}

\newcommand{\beqn}{\begin{eqnarray}}
\newcommand{\eeqn}{\end{eqnarray}}
\newcommand{\be}{\begin{equation}}
\newcommand{\ee}{\end{equation}}
\newcommand{\ba}{\begin{array}}
\newcommand{\ea}{\end{array}}

\newcommand{\bpr}{\begin{proof}}
\newcommand{\epr}{\end{proof}}

\newcommand{\bH}{{\bf H}}

\newcommand{\cA}{{\cal A}}
\newcommand{\bA}{{\bf A}}

\newcommand{\cH}{{\cal H}}
\newcommand{\aH}{{\mathbb H}}

\newcommand{\rS}{{\rm S}}

\newcommand{\aX}{{\mathbb X}}

\newcommand{\bX}{{\bf X}}

\newcommand{\aY}{{\mathbb Y}}

\newcommand{\ci}{\cite}

\newcommand{\De}{\Delta}

\newcommand{\ds}{\displaystyle}
\newcommand{\fr}{\frac}

\newcommand{\ga}{\gamma}

\newcommand{\la}{\label}
\newcommand{\lam}{\lambda}

\newcommand{\na}{\nabla}
\newcommand{  \om}{  \omega}
\newcommand{  \Om}{  \Omega}
\newcommand{\vp}{\varphi}

\newcommand{ \ov}{ \overline}

\newcommand{\pa}{\partial}

\newcommand{\curl}{{\rm curl\5}}

\newcommand{\si}{\sigma}

\newcommand{\ve}{\varepsilon}

\newcommand\C{{\mathbb C}}
\newcommand\R{{\mathbb R}}

\newcommand{\bB}{{\bf B}}

\newcommand{\bP}{{\bf P}}

\newcommand{\vka}{\varkappa}

\newcommand{\Id}{{\rm Id}}

\newcommand{\cm}{{\rm m}}

\newcommand{\5}{{\hspace{0.5mm}}}

\newcommand{\const}{\mathop{\rm const}\nolimits}

\newcommand{\rIm}{{\rm Im\5}}

\newtheorem{theorem}{Theorem}[section]

\newtheorem{definition}[theorem]{Definition}

\newtheorem{lemma}[theorem]{Lemma}
\newtheorem{example}[theorem]{Example}

\newtheorem{remark}[theorem]{Remark}

\newtheorem{remarks}[theorem]{Remarks}
\newtheorem{cor}[theorem]{Corollary}
\newtheorem{proposition}[theorem]{Proposition}

\newcommand{\bd}{\begin{definition}}
 \newcommand{\ed}{\end{definition}}
\newcommand{\bt}{\begin{theorem}}
 \newcommand{\et}{\end{theorem}}
\newcommand{\bqt}{\begin{qtheorem}}
 \newcommand{\eqt}{\end{qtheorem}}

\newcommand{\bp}{\begin{proposition}}
 \newcommand{\ep}{\end{proposition}}

\newcommand{\bl}{\begin{lemma}}
 \newcommand{\el}{\end{lemma}}
\newcommand{\bc}{\begin{cor}}
 \newcommand{\ec}{\end{cor}}

\newcommand{\bex}{\begin{example}}
 \newcommand{\eex}{\end{example}}
 
\newcommand{\bexs}{\begin{examples}}
 \newcommand{\eexs}{\end{examples}}

\newcommand{\bexe}{\begin{exercice}}
 \newcommand{\eexe}{\end{exercice}}

\newcommand{\br}{\begin{remark}}
 \newcommand{\er}{\end{remark}}
 
\newcommand{\brs}{\begin{remarks}}
 \newcommand{\ers}{\end{remarks}}

\newcommand{\bce}{\begin{center}}
\newcommand{\ece}{\end{center}}

\begin{document}
\begin{center}

{\huge On  periodic solutions  
for the Maxwell--Bloch equations}
\footnote{Preprint arXiv: 2312.08180  [quant-ph]}
\bigskip\smallskip

 {\Large A. I. Komech
 }
  \smallskip\\ 
{\it
\centerline{
}
{\it Dobrushin Lab., IITP RAS, Moscow\\
Department Mechanics-Mathematics, Moscow State University}

 }

\centerline{akomech@iitp.ru}

\end{center}

\setcounter{page}{1}
\thispagestyle{empty}

\begin{abstract}

  We consider the Maxwell--Bloch system  which is
      a finite-dimensional
approximation of the coupled nonlinear Maxwell--Schr\"odinger equations.
The approximation consists of one-mode Maxwell field
coupled to $N\ge 1$ two-level molecules.
Our main result is the existence of solutions  with time-periodic Maxwell field.
For the proof 
we construct time-periodic solutions 
to the reduced system with respect to  the symmetry gauge group $U(1)$.
The solutions  correspond to fixed points of the Poincar\'e map,
which are constructed 
using the  contraction of high-amplitude Maxwell field
and  the Lefschetz theorem.
The theorem is applied to a suitable {\it modification} 
of the reduced equations which defines a smooth dynamics 
on the {\it  compactified} phase space.
The crucial role
is played by the fact  that
 the Euler characteristic of the compactified space is strictly greater than the same
of the infinite component. 

  \end{abstract}
  
  \noindent{\it MSC classification}: 
  34C25,
  58D19,
  57Q65,
  	32S50,  	
  78A40, 
  78A60.

  \smallskip
  
    \noindent{\it Keywords}: 
  Maxwell--Bloch equations; laser; attractor; periodic solutions; Rabi solution;  
  gauge group; reduced dynamics; Hopf fibration;   
Poincar\'e map;  fixed points; Lefschetz theorem; Euler characteristic.

\tableofcontents

\setcounter{equation}{0}
\section{Introduction}
The paper addresses an old 
problem of quantum optics on existence of time-periodic solutions 
to the nonlinear coupled Maxwell--Bloch equations (MB)
\ci{SZ1997}.
The equations were introduced by Lamb \ci{L1964}
as a finite-dimensional 
approximation  of the semiclassical  Maxwell--Schr\"odinger system \ci{AE1987,GNS1995,K2022p,Kprep2021}.
The equations
were traditionally used in many works for  semiclassical 
description of the laser action
using perturbation methods and 
different hypotheses on the molecular currents \ci{H1984,KB2003, SSL1978,S1986,S2012}.
However, up to now, there are very few rigorous results.

The main problem of the theory of the laser action is the convergence of the Maxwell field to a high amplitude time-periodic regime in the case
of sufficiently intense pumping. 
In
the
present paper
we establish the existence of solutions with time-periodic Maxwell field
{for the MB equations with $N\ge 1$ two-level molecules
in the case
of time-periodic pumping. For $N=1$,
the equations read as}
 \be\la{HMB2} 
\left\{\ba{rcl}
\dot A(t)\!\!&\!\!=\!\!&\!\!B(t),\quad
\dot B(t)=-\Om^2 A(t)\!-\!\si  B(t)+cj(t)
\\\\
 i\hbar\dot C_{1}(t)\!\!&\!\!=\!\!&\!\!\hbar\om_1C_{1}(t)+ia(t)\,
  C_{2}(t),\quad
i\hbar\dot C_{2}(t)=\hbar \om_2C_{2}(t)-ia(t)\,
  C_{1}(t)
 \ea\right|,\qquad t>0,
\ee
where 
$A(t),B(t)\in\R$ represent the Maxwell filed, while $C_1(t),C_2(t)\in\C$
represent the Schr\"odinger wave function;
$\Om>0$ is the resonance frequency, $\si>0$ is the dissipation coefficient, 
$c$~is the speed of light,  $\hbar$ is the Planck constant, and
 $\hbar\om_2>\hbar\om_1$ are the energy levels of active molecules.
The current $j(t)$ and the function $a(t)$ are given by
\be\la{JOm}
j(t)=
{2}q
\,
 \rIm[\ov C_{1}(t)C_{2}(t)], \quad
a(t)=\fr qc [A(t)+A_p(t)];\qquad q={\om  p},\quad\om=\om_2-\om_1>0,
\ee
where $A_p(t)$ represents an external Maxwell field (the pumping), and $p>0$
is proportional to the molecular dipole moment of the molecule; see Appendix \ref{aA}.

The {\it charge conservation} holds: $|C_{1}(t)|^2+|C_{2}(t)|^2=\const$ for  $t\in\R$.
The conservation follows by differentiation from the last two equations of  (\ref{HMB2})
since the function $a(t)$ is real-valued.
We will consider solutions with the $\const=1$, so
\be\la{Blcc}
|C_{1}(t)|^2+|C_{2}(t)|^2=1,\qquad t\ge 0.
\ee
Accordingly, the phase space of the MB system (\ref{HMB2}) is $\aX=\R^2\times \rS^3$.
We assume that
$
A^p(t)\in C(\ov\R^+).
$
Then the existence and uniqueness  of  global solutions 
$X(t)\in C(\ov\R^+,\aX)$
to the  system (\ref{HMB2}) with any given initial state $X(0)\in \aX$
follows from 
  the \emph{a priori} bound 
   \be\la{apri}
|X(t)|\le D_1|X(0)|e^{-\ga t}+D_2,\quad t>0;\qquad D_1, D_2, \ga>0,
\ee
which is proved in Section \ref{swp}.

Our main result is the existence 
of solutions with time-periodic Maxwell field
for the MB equations 
(\ref{HMB2}), (\ref{HMB3N})
with  any number of molecules $N\ge 1$ in the case of time-periodic pumping:
\be\la{pum}
A_p(t+T)=A_p(t),\quad  t\in\R;\qquad T=2\pi/\Om_p,
\ee
where $\Om_p$ is the pumping frequency. 
\smallskip

Let us comment on  our approach.
The equations (\ref{HMB2}) are invariant with respect to the 
group $G=U(1)$ of
gauge transformations
\be\la{gt}
g(\theta)
(A, B,C)=
(A, B, e^{i\theta}C),\qquad \theta\in [0,2\pi];
\qquad A,B\in\R,
\quad C=(C_1,C_2)\in\C^2.
\ee
The action  (\ref{gt})
does not affect the Maxwell field. Hence,
it suffices to prove  the existence of $T$-periodic solutions
 for the MB equations 
{\it reduced by the symmetry gauge group} $U(1)$; see (\ref{gt}), (\ref{HMB2f}).

  \br\rm For the corresponding solutions 
to the MB equations, the  Maxwell field, the current, and the
population  inversion
are time-periodic, while   
the wave function can acquire a unitary factor in 
each
period; see 
(\ref{perh}) and (\ref{perh2}).
\er

  The time-periodic solutions correspond to fixed points 
  of the Poincar\'e map for the reduced equations.
Our construction of  the fixed points 
relies on the 
{\it contraction} (\ref{apri}) of 
 high-amplitude  Maxwell field
and  the Lefschetz theorem on fixed points and Euler characteristic
  \ci{GP1974}.
  The main issue in our 
  application of the theorem 
 is
 that it is known for compact spaces,
while for the MB equations, the phase space {$\aX$} is not compact.
Accordingly, 
the intersection number can be changed in the homotopy if  
some of the fixed points run off to infinity in finite time.
To prevent such runaway, we use
the contraction (\ref{apri})
for the \emph{a priori} bound of the fixed points, thus
reducing the problem to the compact space.

The Lefschetz theorem is applied to a suitable {\it modification} 
of the reduced equations which defines a smooth dynamics 
on the {\it compactified} phase space with a ``repulsion" from infinity.
The crucial role
is played by the fact  that
 the Euler characteristic of the compactified space is strictly greater than the same
of the infinite component; see 
Remark~\ref{rred}.

We give detailed proofs for the 
MB equations with one active molecule
described by the two-level Schr\"odinger equation.
Further we extend the result to the case of any number of the molecules $N> 1$
which is typical for the laser action with $N\sim 10^{20}$; see \ci{K2022p}.

As a by-product, we obtain
for the MB equations
 the existence 
of a compact global attractor \ci{CV2000,T1997}. 
\smallskip

Let us comment on previous related results.
  The problem of existence of time-periodic solutions 
has been
 discussed since 1960s. 
 The first results in this direction were obtained recently in
  \ci{CLN2017}-\ci{CLV2019} and \ci{WWG2018}
 for various versions
 of the MB equations. In  \ci{WWG2018},
  the approximate  $N$th order time-periodic
  solutions 
  were constructed by perturbation techniques.
   For the phenomenological model \ci{A1985,AB1965},
  time-periodic solutions
  were constructed in \ci{CLV2019} in the absence of
   time-periodic pumping \ci[Eq. (1)]{CLV2019}  for small interaction constant.
   The solutions are constructed as the result of a bifurcation
   relying on  homotopy invariance of the degree  \ci{CLN2017} and develops the
   averaging arguments  \ci{CL2018}. 
   {The period is determined by bifurcation, and is not related to 
    an external source}.
   
   Up to our knowledge, periodic solutions for the MB equations 
   without the smallness assumptions
   were not constructed until now.

   \br\la{rperi}\rm
i)   By Lemma \ref{lperh}, 
for solutions of the nonreduced Maxwell--Bloch system,
corresponding to the periodic solutions of the reduced equations, 
 the Maxwell field, current, and the population inversion are time-periodic,
 while the wave function can acquire a unitary factor in each period (\ref{perh}).
  So the time-periodic solutions to the nonreduced Maxwell--Bloch system
 with nonzero pumping
 probably do not exist, see Remark~\ref{rperh}. 
   \smallskip\\
 ii)
In  \ci{CLN2017}-\ci{CLV2019} and \ci{WWG2018},
 the time-periodic solutions are constructed 
 for the MB systems with wave function  replaced by
the
 density matrix which is
gauge-invariant, so the reduction is not needed.

 \er
 The contraction of high-amplitude  Maxwell fields
and well-posedness are proved in Section \ref{swp}.
In Section \ref{sU}, we introduce the reduced dynamics.
In Section \ref{sper1} solutions with time-periodic Maxwell field
are constructed for the equations with one molecule, and 
in Section \ref{sperN} -- for  equations with many molecules.
In Appendix \ref{aA} we recall the introduction of the MB equations.

\setcounter{equation}{0}
\section{The contraction of high-amplitude Maxwell field and well-posedness}
\la{swp}
In this section, we prove the a priori bound (\ref{apri})
assuming that $A^e(t)\in C[0,\infty)$.  The bound implies
 the well-posedness of the MB system (\ref{HMB2}) in
the phase space $\aX=\R^2\times \rS^3$.
 The Schr\"odinger amplitudes $C_1(t),\,C_2(t)$
are bounded by the charge conservation (\ref{Blcc}).  Hence, it remains to prove 
the \emph{a priori} estimates for the Maxwell amplitudes $(A(t),B(t))$.   
\bl
Let $A^e(t)\in C[0,\infty)$. Then there exists the Lyapunov function
$V(A,B)$ such that 
\be\la{aprL}
a_1[A^2+B^2]\le V(A,B)\le a_1[A^2+B^2]\quad {\rm where}\quad a_{1,2}>0,
\ee
 and for solutions to (\ref{HMB2}),
\be\la{Vder3}
\fr d{dt} V(A(t),B(t))\le -\ga
V(A(t),B(t))
+D,\qquad t>0;
\quad \ga,D>0.
\ee
 \el
\bpr
Denote by $E(A,B)=\fr1{2}(\Om^2A^2+B^2)$  ``the energy" of the Maxwell field.
The first two equations of (\ref{HMB2}) imply for 
\be\la{denE}
\fr d{dt}E(A(t),B(t))=\Om^2A(t)B(t)
+B(t)(-\Om^2A(t)-\si B(t)+cj(t)
=
-\si B^2(t)+ cB(t) {j(t)}.
\ee
We introduce the Lyapunov function  following
the standard approach 
to dissipative perturbations of the Hamiltonian systems
 \ci{BV1992,H1981}:
$
V(A,B)=E(A,B)+\ve AB,
$
where sufficiently small $\ve>0$ will be chosen below.
Differentiating, we obtain for $V(t)=V(A(t),B(t))$:
\beqn\nonumber
\dot V(t) &=&\dot E(t)+\ve\dot A(t)B(t)+\ve A(t)\dot B(t)\\
\nonumber
&=&-\si B^2(t) +cB(t){j(t)} +\ve B^2(t)+\ve A(t)[-\Om^2 A(t)-\si  B(t)+c j(t)]\\
\nonumber
&=&-(\si-\ve) B^2(t)-\ve\Om^2 A^2(t)-\ve\si A(t) B(t)+c{j(t)}(\ve A(t)+B(t)).
\eeqn
Note that $|\ve\si AB|\le\fr\si{2}B^2+\fr12\si\ve^2 A^2$.
Hence,   for small $\ve>0$, 
\be\la{Vder2}
\dot V(t)\le-[\fr{\si}{2}-\ve]B^2(t)-\fr\ve2 \Om^2A^2(t)+c{j(t)}(\ve A(t)+B(t)),\qquad t>0.
\ee
Moreover,  $|j(t)|\le 2\vka $ by (\ref{JOm}) and (\ref{Blcc}). 
Hence, for  sufficiently small $\ve>0$, we get (\ref{Vder3}).
\epr

\bc
Solving the inequality (\ref{Vder3}) in the region $V>D/\ga $, we obtain:
\be\la{Vdec}
V(t)\le 
V(0)e^{-\ga  t}+ \fr{D}\ga  {(1- e^{-\ga  t})},\qquad t>0.
\ee
Hence,
for solutions to (\ref{HMB2}),
 the following bounds hold:
 \be\la{apr2}
|A(t)|^2+|B(t)|^2\le d_1(|A(0)|^2+|B(0)|^2)e^{-\ga t}+d_2,\,\,\,\, t>0;
\quad d_1, d_2, \ga>0.
\ee
\ec
\br\rm
i)
{The bound (\ref{apr2}) trivially follows by solving  the first two equations of 
(\ref{HMB2}). However, it does not imply 
 the bound (\ref{Vder3}) which plays 
 crucial
 role in our construction of the time-periodic solutions.
\smallskip\\
ii) The Lyapunov function $V(A,B)$ is the quadratic form with the level lines
which are ellipses transversal  to the vector field  $(B,-\Om^2A-\si B)$.
In particular, the transversality does not hold for $V(A,B)=E(A,B)$ as is seen
from (\ref{Vder2}) with $\ve=0$.
}
\er

\br\rm
The bound (\ref{apri}) implies that
there exists  the compact global attractor  
for the MB equations which is
the minimal  subset $\cA\subset\aX\,\,$ such that
$$
X(t)\to \cA,\qquad t\to\infty
$$
for any solution $X(t)$ to (\ref{HMB2}). 
This follows  by the standard techniques \ci{BV1992,CV2000,KK2020,T1997}.
\er

\setcounter{equation}{0}
\section{Gauge symmetry and reduced dynamics in the Hopf fibration}\la{sU}
Recall that 
the phase space of the MB system 
is
$\aX=\R^2\times \rS^3$ due to
the charge conservation (\ref{Blcc}).
The MB  system (\ref{HMB2}) 
is
 $U(1)$-invariant with respect
to the action (\ref{gt}).
This means that the function $g(\theta)X(t)$ is a solution if $X(t)$ is. This is obvious
from
(\ref{HMB2}) and (\ref{JOm}).
  Let us denote the factorspace $\aY=\aX/U(1)$
  which is the space of all orbits $X_*=\{g(\theta)X:\theta\in[0,2\pi]\}$ with $X\in\aX$.
  Denote the map 
  \be\la{Hr}
  \Pi:X=(A,B,C)\mapsto X_*=(A,B,C_*)\in\aY,\qquad X\in\aX.
  \ee
  The factorspace  $\aY$ is the smooth manifold $\R^2\times \rS^2$ since 
the map $h:C\mapsto C_*$ is the Hopf fibration $\rS^3\to \rS^2$.
The MB system (\ref{HMB2}) induces the corresponding reduced dynamics in the factorspace
 $\aY$ which can be written as
\be\la{HMB2f}
\dot Y(t)=F(Y(t),t),\qquad t\ge 0,
\ee
where $F(\cdot, t)$ is the smooth vector field on $\aY$ which continuously depends on $t\ge 0$.
\bl\la{lwpr}
   The reduced equation (\ref{HMB2f}) admits a unique global solution
   $Y(t)=(A(t),B(t), C_*(t))$ for every initial state $Y_0\in\aY$.
\el
\bpr
The uniqueness follows from the smoothness and continuity of $F(\cdot, t)$.
To prove the existence, take any point $\hat Y_0\in \Pi^{-1}Y(0)$ and set
$Y(t)=\Pi \hat Y(t)$, where $\hat  Y(t)$ is the solution to the MB equations (\ref{HMB2}) with the initial state $\hat Y_0$. 
Then $Y(t)$ is the solution to (\ref{HMB2f}) by definition of the reduced dynamics,
and the initial state $Y(0)=\Pi\hat Y(0)=Y_0$.
\epr
For $ X=(A,B, C)\in\aX$,
the  current $j=2q\,\rIm[\ov C_1C_2]$ 
and the population inversion $I:=|C_2|^2-|C_1|^2$
are invariant with respect to the gauge transformations (\ref{gt}).
So, $j_*(h\,C):=j(C)$ is a bounded  smooth  function on $\rS^2$.
In particular,  the first line of (\ref{HMB2})
can be written as
\be\la{ABw}
\dot A(t)= B(t),\quad
\dot B(t)=-\Om^2 A(t)\!-\!\si  B(t)+cj_*(C_*(t)).
\ee
\bl\la{lperh}
 Let  $Y(t)=(A(t),B(t),C_*(t))$ be a $T$-periodic solutions 
 to the reduced dynamics (\ref{HMB2f}),
 and $\hat Y(t)=(\hat A(t),\hat B(t),\hat C(t))$ be the solution
 to 
 the nonreduced  MB equations (\ref{HMB2})
 with any $\hat Y(0)\in \Pi^{-1}Y(0)$.
 Then
 the Maxwell amplitudes $(\hat A(t),\hat B(t))$, the current 
 $j(t)=q\rIm[\ov {\hat C_1}\hat C_2(t)]$,
 and the population inversion $I(t):=|\hat C_2(t)|^2-|\hat C_1(t)|^2$
 are $T$-periodic, 
 while the wave function  can acquire a unitary factor:
 \be\la{perh}
 \hat C(t+T)=e^{i\theta(t)}\hat C(t),\qquad t\ge 0,\quad \theta(t)\in[0,2\pi].
 \ee
\el
\bpr
By definition of the reduced dynamics, $\Pi\hat Y(t)=Y(t)$, that is 
$(\hat A(t),\hat B(t),h\hat C_*(t))
=(A(t), B(t), C(t))
$. Hence,
the Maxwell amplitudes $(\hat A(t),\hat B(t))$ are $T$-periodic. 
Similarly,  
the Hopf projection  $h \hat C(t)$ is also $T$-periodic, that is equivalent to 
(\ref{perh}). Finally, (\ref{perh}) impies that  $I(t)$ is $T$-periodic.
\epr

\br\la{rperh}\rm
 The relation (\ref{perh}) suggests that
  for the nonreduced system (\ref{HMB2}) with  $T$-periodic pumping (\ref{pum}),  $T$-periodic solutions
 probably do not exist. 
     This conjecture is confirmed by the Rabi solution \ci{R1937} since it can
 contain incommensurable  frequencies \ci[(4.8)]{GK2004}.
 Our topological arguments also do not work for the nonreduced system;
 see Remark \ref{rred}.
 
 \er


\setcounter{equation}{0}
\section{Periodic solutions for the reduced MB equations with one particle}
\la{sper1}
In this section, we consider the reduced dynamics (\ref{HMB2f}) 
with a time-periodic  pumping (\ref{pum}).
 Then  (\ref{HMB2f})  is a time-periodic system:
 \be\la{HMB2p}
F(Y,t+T)=F(Y,t),\qquad  Y\in \aY,\qquad  t\ge 0.
\ee
 The main result of present paper is the following theorem.

 \bt\la{tper}
 Let 
 (\ref{pum})  hold. 
 Then the MB equations  (\ref{HMB2}) admit
 solutions with
  $T$-periodic Maxwell field:
  \be\la{Y1T}
 (A(t+T),B(t+T))=(A(t),B(t)),\qquad t\ge 0.
 \ee
 \et

To prove Theorem \ref{tper},
 it suffices to construct 
 $T$-periodic solutions $Y(t)$ for
  the reduced dynamics (\ref{HMB2f}):
 \be\la{YT}
 Y(t+T)=Y(t),\qquad t\ge 0.
 \ee
  
  \subsection{A priori estimate for fixed points of the Poincar\'e map}
   Solutions to  (\ref{HMB2f}) admit the representation 
  \be\la{Ut}
  Y(t)=U(t)Y(0),\qquad t\ge 0,
  \ee
  where $U(t):\aY\to \aY$ is the diffeomorphism and $U(0)=\Id$ is the identity map.
 The map
 $U(T)$ is homotopic to the identity since
 $U(t)$
  depends
 continuously on $t\in\R$. 
 The existence of $T$-periodic solution (\ref{YT})
  is equivalent to the fact that
the Poincar\'e map $U(T)$ admits at least one fixed point.
Let us denote the set of all fixed points by
\be\la{svfp}
\Phi=\{Y_\#\in \aY: U(T)Y_\#=Y_\#
\}.
\ee
\bl\la{lfp}
The set $\Phi$ is bounded in $\aY$.
\el
 \bpr
 Denote by $v$
 the vector field on $\R^2$ corresponding  to the equations (\ref{ABw}):
 for $M=(A,B)\in\R^2$ and $C_*\in \rS^2$,
\be\la{vfi}
v(M,C_*)=\left(\!\!\!
\ba{c}B \\
-\Om^2A-\si B+cj_*(C_*)\ea\!\!\!
\right),\qquad j_*(C_*)=2q\,\rIm [\ov C_1(t)C_2(t)].
\ee
By (\ref{Vder3}) and (\ref{aprL}),
for large $|M|$,
the field $v(M,C_*)$ is directed ``towards the origin":
\be\la{vdir}
v(M,C_*)\cdot \na V(M)\le -a_1\ga |M|^2+D<0,\qquad |M|>R(D).
\ee
Therefore, the region $|M|>R(D)$ does not contain fixed points.
\epr

\subsection{Modified dynamics on the compactified phase space}

   We are going to apply 
 the Lefschetz theorem \ci[p. 120]{GP1974}:
  the number of fixed points (counted with multiplicities) of any continuous map
 homotopic to the identity map
 of a compact space to itself 
 is equal to the Euler characteristic  of this space.
 However,
 in our case the phase space $\aY=\R^2\times \rS^2$ is not compact, so 
 we first have to reduce the question to a compact case.
 
 Introduce the {\it compactification} $\aY_c=\rS^2_c\times \rS^2$ of the phase space,
 where
 $\rS^2_c=\R^2\cup *$ 
 with the neighborhoods of the ``infinite point" $*$ defined by
 \be\la{neib}
 O_R(*)=\{M\in\R^2:|M|>R\},\qquad R>0.
  \ee
  Let us define a {\it modification} of  dynamics (\ref{HMB2f}) on the space $\aY_c$
  by the system 
  \be\la{redm}
  \dot Y_c(t)=F_c(Y_c(t),t),\qquad t > 0.
  \ee
   The goal of the modification is to   obtain a smooth vector field $F_c(\cdot,t)$ on the compactification $\aY_c$ and 
   to keep the same set of all fixed points in $\aY$.
   Denote $M(t)=(A(t),B(t))$ and first modify the system (\ref{HMB2}) by
    \be\la{HMB2m} 
    \left\{\ba{rcl}
\dot M(t)&=&v_R(M(t), C_1(t),C_2(t))
\\\\
 i\hbar\dot C_{1}(t)&=&\zeta_R(|M|)[\hbar\om_1C_{1}(t)\!+\!ia(t)\,
  C_{2}(t)]\\\\
i\hbar\dot C_{2}(t)&=&\zeta_R(|M|)[\hbar \om_2C_{2}(t)\!-\!ia(t)\,
  C_{1}(t)]
  \ea\right|,\qquad t>0,
\ee
where
\be\la{ac}
 \zeta\in C^\infty[0,\infty),\qquad
 \zeta_R(r)=  \left\{\ba{ll} 1,& r\le R\\0,& r\ge R+1\ea\right|.
 \ee
 For  the modified vector field $v_R$ we require
 \be\la{vdir2}
 v_R(M,C_1,C_2)=v(M,C_1,C_2),\quad |M|<R,
 \quad\ {\rm and}\ \quad
v_R(M,C_1,C_2)\cdot \na V(M)<0,\quad |M|>R.
\ee
 By (\ref{vdir}),
such modification exists for $R>R(D)$.
Similarly to (\ref{HMB2}),
 the modified system (\ref{HMB2m})
 is invariant with respect to the action (\ref{gt}).
 Accordingly,
define the modified dynamics (\ref{redm}) as the corresponding  reduction of  (\ref{HMB2m}).
Moreover, we can require that for large $M$ the modified field is radial
and  
\be\la{vdir3}
v_R(M,C_1,C_2)=-M/|M|^2,\qquad |M|\ge R_c\ge R.
\ee
Then the vector field 
of the modified system (\ref{redm}) 
 is smooth on the compactified space $\aY_c=\rS^2_c\times \rS^2$ with the smooth structure 
defined by the coordinates $(M/|M|^2,C_*)$ in a neighborhood of the infinite sphere
$\rS^2_*=*\times \rS^2$.

As a result, we have proved the following lemma.

\bl There exists an $R>0$ and   a  modification (\ref{redm}) of the reduced system (\ref{HMB2p})
with the following properties:
\smallskip\\
i)  The modified system (\ref{redm}) is smooth on 
the compactified space $\aY_c=\rS^2_c\times \rS^2$.
\smallskip\\
ii) For both systems (\ref{redm}) and (\ref{HMB2p}), the fixed points
of the Poincar\'e map  with $|M|<R$  are identical;
\smallskip\\
iii) For both systems, the fixed points of the Poincar\'e map  with $|M|>R$  do not exist;
\smallskip\\
iv) The identity (\ref{vdir3}) holds for the modified system.
\el

Now we can apply the Lefschetz theorem to the Poincar\'e map $U_c(T):\aY_c\to \aY_c$
which corresponds to the modified system (\ref{redm}).
Indeed,
the map is homotopic to the 
identity map,
and the Euler characteristic
is given by  $\chi(\aY_c)=\chi(\rS^2_c)\chi(\rS^2)=4$. 
Hence, $U_c(T)$ admits four fixed points in $\aY_c$
counted with their multiplicities.
Further, by (\ref{vdir3}),
the infinite sphere $\rS^2_*$ is 
invariant under $U_c(T)$, and the number of fixed points $Y_\#\in \rS^2_*$ with multiplicities equals
 the Euler characteristic 
 $\chi(\rS^2_*)=2$.
Hence, $U_c(T)$ has at least one fixed point  $Y_\#\in \aY=\aY_c\setminus \rS^2_*$.
Finally, $Y_\#$ is also the fixed point for $U(T)$.
Theorem \ref{tper} is proved.

  \br\la{rred}\rm
  The Lefschetz theorem gives the conclusion
  due to the bound (\ref{apr2}) for the reduced Maxwell--Bloch equations
  with the phase space $\aY=\R^2\times \rS^2$.
  Note that the similar bound (\ref{apri}) holds for the nonreduced system (\ref{HMB2})
  with the phase space $\R^2\times \rS^3$,
  however in this case 
  the Lefschetz theorem does not give the desired result since the Euler characteristic is
  $\chi(\aX)=\chi(\rS^3)=0$.
This is what made necessary
    the reduction of the MB system by the symmetry group $U(1)$.

  \er

\br\la{rsdv}\rm
The crucial role in our proof is played by the inequality (\ref{vdir})
which means that for large $|M|$, the vector field $v(M,C_*)$ admits the Lyapunov function
$V(M)\to\infty$ as $|M|\to\infty$.
Such function does not exist,
for example,
 for the vector field of the system $\dot A=\dot B=1$, $\dot C=0$.
Accordingly, the Poincar\'e map is the shift $(A,B,C)\mapsto (A+T,B+T,C)$,
so the system does not admit solutions with $T$-periodic $A(t)$ and $B(t)$.

\er



\setcounter{equation}{0}
\section{Maxwell--Bloch equations with many molecules}\la{sperN}
Let us write 
the MB equations for many molecules.
 Now  (\ref{solMB2}) becomes
 \be\la{solMB2N}
\psi_n(x,t)=C_{n,1}(t)\vp_{n,1}(x)+C_{n,2}(t)\vp_{n,2}(x),\qquad x\in V,
\quad 
n\in\ov N:=\{1, \dots, N\}.
\ee
 The MB system reads, similarly to (\ref{HMB2}), as 
 \be\la{HMB3N} 
\left\{\ba{rcl}
\dot A(t)\!\!&\!\!=\!\!&\!\! B(t),\quad
\dot B(t)=-\Om^2 A(t)\!-\!\si  B(t)+cj(t)
\\\\
 i\hbar\dot C_{n,1}(t)\!\!&\!\!=\!\!&\!\!\hbar\om_1C_{n,1}(t)+ia(t)\,
  C_{n,2}(t),\quad
i\hbar\dot C_{n,2}(t)=\hbar \om_2C_{n,2}(t)-ia(t)\,
  C_{n,1}(t),\quad n\in\ov N
 \ea\right|,
\ee
where the current $j(t)$ is given by
\be\la{JOm2N}
j(t)=
\sum_{n=1}^N\vka_n
 \rIm[\ov C_{n,1}(t)C_{n,2}(t)].
\ee

All the results above 
for the system  (\ref{HMB2}) 
can be extended to the system  (\ref{HMB3N}). In particular, 
the charge conservation (\ref{Blcc}) holds for each active molecule:
\be\la{BlccN}
|C_{n,1}(t)|^2+|C_{n,2}(t)|^2=1,\qquad t\ge 0, \quad n\in\ov N.
\ee
The gauge group $G=[U(1)]^N$
acts on the phase space $\aX=\R^2\oplus \C^{2N}$ by
\be\la{actG2}
g(e^{i\theta_1},\dots, e^{i\theta_{N}})
(A, B,C)=
(A, B, e^{i\theta_1}C_1,\dots,  
e^{i\theta_{N}}C_N)\qquad C(t)=(C_1(t),\dots,C_N(t))\in [\rS^3]^{N},
\ee
where $C_n(t)=( C_{n,1}(t),\,C_{n,2}(t))\in \rS^3$.
This action commutes with the dynamics (\ref{HMB3N}),
hence the latter induces the corresponding  reduced dynamics   on the factorspace
$\aY=\aX/G=\R^2\times [\rS^2]^N$. The action  (\ref{actG2})
does not affect the Maxwell field. 
 \bt\la{tperN}
 Let   (\ref{pum}) hold. 
 Then the MB equations 
 (\ref{HMB3N}) admit solutions with $T$-periodic Maxwell field.
 \et
 The proof of the theorem relies on a minor modification of constructions and arguments used above
in the case $N=1$.
 The relation (\ref{perh}) holds as in the case $N=1$ 
with the obvious modification: now
 \be\la{perh2}
 C(t+T)=g(e^{i\theta_1(t)},\dots, e^{i\theta_{N}(t)})C(t),\qquad t\ge 0,\quad \theta_k(t)\in[0,2\pi].
 \ee

\br\la{reson}\rm
Our assumption (\ref{pum})   guarantees
the existence of solutions
to the MB equations
with time-periodic 
Maxwell field.
However,
we do not impose the resonance conditions
\be\la{reso}
\Om_p\approx\Om\approx\om,
\ee
which physically are responsible for the  ``lasing", that is, the effective laser  action.
\er

\section{Acknowledgements} 
The author thanks V.A. Vassiliev for 
helpful discussions of topological aspects, and 
S. Kuksin, M.I. Petelin, A. Shnirelman and H. Spohn
 for longterm fruitful discussions. The research was carried out within the state assignment of Ministry of Science and Higher Education of the Russian Federation for IITP RAS.


\appendix 

\section{The Maxwell--Bloch equations as the Galerkin approximation}
\la{aA}
Here we introduce the MB equations (\ref{HMB2})
as the Galerkin approximation of the semiclassical Maxwell--Schr\"odinger equations studied in \ci{BT2009,GNS1995,S2006} (see also \ci{K2019phys,K2013,K2022,KK2020}) and endowed
with the damping and pumping.
The MB equations 
describe the coupling of
  one-mode Maxwell field
 with a two-level molecule in a bounded cavity $V\subset\R^3$:
 \be\la{solMB2}
\bA(x,t)=A(t)\bX(x),\quad
\psi(x,t)=C_{1}(t)\vp_{1}(x)+C_{2}(t)\vp_{2}(x),\qquad x\in V.
\ee
 Here $\bX(x)$ and $\vp_{l}$ are suitable {\it normalized} eigenfunctions of the Laplace
and Schr\"odinger operators under suitable boundary value conditions:
\be\la{eigem}
\De \, \bX (x)=- \fr{\Om ^ 2}{c^2} \bX (x), \quad x \in V;
\qquad \ds \bH\vp_{l}(x)=\hbar\om_{l}\vp_{l}(x),\quad
 x\in V,\quad l=1,\,2.
\ee
We introduce the Schr\"odinger operator
$\bH:=-\fr{\hbar^2}{2\cm}\De  + e\Phi (x)$, where $\Phi (x)$
is the molecular (ion's) potential.
  The semiclassical
MB equations 
can be defined as the Hamiltonian equations with dissipation:
\be\la{HMB} 
\fr1{c^2}\dot A(t)=\pa_B H,\quad \fr1{c^2}\dot B(t) =-\pa_A H-\fr\si{c^2} B;
\qquad i\hbar\dot C_{l}(t)=\pa_{\ov C_{l}}H,\quad l=1,\,2,
\ee
where $\si>0$ is the electrical conductivity of the cavity medium.
The Hamiltonian is defined as 
\be\la{HH}
H(A,B,C,t)=\cH(A\bX,
B\bX, C_{1}\vp_{1}
+C_{2}\vp_{2},t),\qquad C=(C_1,C_2),
\ee 
where 
$\cH$ is the Hamiltonian of the 
Maxwell--Schr\"odinger equations 
with pumping.
We neglect the spin and scalar potential which can be easily added. 
In the Heaviside--Lorentz units \ci{Jackson} (which are also 
{\it  unrationalized Gaussian units}),
the Hamiltonian $\cH$ 
reads as
\be\label{enc}
\cH(\bA,\bB,\psi,t)=
\fr12[\Vert \fr1c \bB\Vert^2
+\Vert \curl\bA\Vert^2]+\langle\psi, \aH(\bA,t)\psi\rangle,
\ee
where
$\Vert\cdot\Vert$ stands for the norm in the phase Hilbert space
$L^2(V)\otimes\R^3$ and
the brackets
$\langle\cdot,\cdot\rangle$ stand for the Hermitian  inner product in
$L^2(V)\otimes\C$.
The  Schr\"odinger operator reads as 
\beqn\la{Sope}
\aH(\bA,t)\!\!&\!\!:=\!\!&\!\!
\fr 1 {2m} [- i \hbar \na- \ds \frac ec (\bA ( x)+\bA_p ( x,t))]^2
+e\Phi(x)
\\\nonumber
\!\!&\!\!=\!\!&\!\! \ds \bH
+\fr {e\hbar} {2m c} \Big[(\bA (x)+\bA_p ( x,t))\circ i\na
+i\na\circ(\bA (x)+\bA_p ( x,t))
\Big]
+\frac {e^2}{2mc^2} (\bA ( x)+\bA_p ( x,t))^2,
\eeqn
where $\bA_p ( x,t)=\bX(x)A_p(t)$ is the pumping.
Substituting (\ref{solMB2}) into (\ref{enc}), we find:
\be\la{Hc}
H(A,B,C,t)=\fr1{2c^2}[B^2
+ \Om^2A^2]+\langle\psi, \aH(\bA,t)\psi\rangle.
\ee
Using (\ref{Sope}), we obtain:
\begin{align}\la{Hc2}
\langle\psi, \aH(\bA,t)\psi\rangle
&=\hbar\om_1|C_{1}|^2+\hbar\om_2|C_{2}|^2
\nonumber\\
&
+i\fr {e\hbar} {2m c}
(A+A_p (t))
\sum_{l,l'}\ov C_{l}C_{l'}
\Big[\langle \vp_{l}(x)\bX (x), \na \vp_{l'}(x)\rangle+
\langle \vp_{l}(x),\na(\bX (x)\vp_{l'}(x))\rangle
\Big]
\nonumber\\
&
+\frac {e^2}{2mc^2} (\bA ( x)
+\bA_p ( x,t))^2.\qquad\qquad\qquad\qquad
\nonumber
\end{align}
Substituting into (\ref{Hc}), 
we get:
\beqn\la{Hc3}
H(A,B,C,t)&=&\fr1{2c^2}[B^2
+ \Om^2A^2]+\hbar\om_1 |C_{1}|^2+
\hbar\om_2|C_{2}|^2+i\fr {\hbar} {2m c}(A+A_p (t))\sum_{l,l'} \ov C_{l}C_{l'} P_{l,l'}
 \nonumber\\
\nonumber\\
&&
+\frac {e^2}{2mc^2} (\bA ( x)+\bA_p ( x,t))^2,\qquad\qquad
\eeqn
where
\be\la{Qll}
P_{l,l'}=e\Big[\langle\bX (x)\vp_{l}(x), \na \vp_{l'}(x)\rangle-
\langle\na\vp_{l}(x),  \bX (x)\vp_{l'}(x)\rangle
\Big].
\ee
 The last term
on the right hand side of (\ref{Hc3}) is negligible 
compared to the first term
because usually $\Om^2\gg \fr{e^2}m$. 
For example, $\Om\approx 3\times 10^{15} s^{-1}$ for the Ruby laser
\ci{KB2003,S2012}, while $\fr{e^2}m\approx 3\times 10^{8}s^{-1}$.
This is why the last term is traditionally neglected \ci[Eq. (44.13)]{Schiff1955};
we will also neglect this term in the Hamiltonian. 
Moreover, we will use the standard
{\it dipole approximation} which physically means that the wavelength $\lam=2\pi c/\Om$
is negligible with respect to the size of a molecule. In this case, 
\be\la{Qll2}
P_{l,l'}\approx 
P_{l,l'}^d
=2e\bX (x_*)\langle\vp_{l}(x), \na \vp_{l'}(x)\rangle,
%
\qquad P_{l,l'}^d=-\ov P_{l,l'}^d,
\ee
where $x_*\in V$ is the  location of the molecule.
As a result, we take the Hamiltonian
in the form
\beqn\la{Hc32}
H(A,B,C,t)=\fr1{2c^2}[B^2
+ \Om^2A^2]+\hbar\om_1 |C_{1}|^2+
\hbar\om_2|C_{2}|^2
 +i\fr {e\hbar} {2m c}(A+A_p (t))\sum_{l,l'}  \ov C_{l}C_{l'}P_{l,l'}^d.\qquad\qquad
\eeqn
The commutation $[\bH,x]=-\fr{\hbar^2}m \na$ implies
the well-known identity \ci[Eq. (44.20)]{Schiff1955}
\be\la{PP}
e\langle\vp_{l}, \na \vp_{l'}\rangle=-\fr {em}{\hbar^2}\langle\vp_{l}, [\bH,x]\vp_{l'}\rangle
=-\fr {em}{\hbar}[\om_l-\om_{l'}]\langle\vp_{l}, x\vp_{l'}\rangle
=-\fr m{\hbar}\De_{ll'}\bP^{ll'},\quad \om_{ll'}:=\om_l-\om_{l'}.
\ee
We have $\bP^{12}=\bP^{21}=\bP=e\langle\vp_{1}, x\vp_{2}\rangle\in\R^3$  since we can assume that
both wave functions $\vp_l$ are real;
here, $\bP$
is the  dipole moment (or polarization) of the molecule
(this explains the term ``dipole approximation").
Hence, (\ref{Hc32}) and (\ref{Qll2}) give 
\be\la{Hc33}
H(A,B,C,t)=\fr1{2c^2}[B^2
+ \Om^2A^2]+\hbar\om_1 |C_{1}|^2+
\hbar\om_2|C_{2}|^2
  -\fr {2\om p}c
p[A+A_p(t)]\,
 \rIm[\ov C_{1}C_{2}],
\ee
where $p=\bX(x_*)\bP$ and $\om=\om_{21}$.
 Now the Hamilton equations (\ref{HMB}) read as (\ref{HMB2}).

\end{document}